\documentclass[aps,prb,twocolumn,amsmath,amssymb,superscriptaddress,floatfix]{revtex4-2}
\usepackage{graphicx} 
\usepackage{bm}
\usepackage[usenames]{color}
\bibstyle{apsrev.bib}
\usepackage{amsmath}

\newcommand{\be}{\begin{equation}}
\newcommand{\ee}{\end{equation}}
\newcommand{\beqn}{\begin{eqnarray}}
\newcommand{\eeqn}{\end{eqnarray}}

\begin{document}

\title{Reducing defect production in random transverse-field Ising chains by inhomogeneous driving fields}

\author{R\'obert Juh\'asz}
\email{juhasz.robert@wigner.hun-ren.hu}
\affiliation{HUN-REN Wigner Research Centre for Physics, H-1525 Budapest, P.O.Box 49, Hungary}

\author{Gergő Roósz}
\email{roosz.gergo@wigner.hun-ren.hu}
\affiliation{HUN-REN Wigner Research Centre for Physics, H-1525 Budapest, P.O.Box 49, Hungary}

\date{\today}

\begin{abstract}
In transverse-field Ising models, disorder in the couplings gives rise to a drastic reduction of the critical energy gap and, accordingly, to an unfavorable, slower-than-algebraic scaling of the density of defects produced when the system is driven through its quantum critical point.  
By applying Kibble-Zurek theory and numerical calculations, we demonstrate in the one-dimensional model that the scaling of defect density with annealing time can be made algebraic by balancing the coupling disorder with suitably chosen inhomogeneous driving fields.
Depending on the tail of the coupling distribution at zero, balancing can be either perfect, leading to the well-known inverse-square law of the homogeneous system, or partial, still resulting in an algebraic decrease but with a smaller, non-universal exponent.
We also study defect production during an environment-temperature quench of the open variant of the model in which the system is slowly cooled down to its quantum critical point. According to our scaling and numerical results, uncorrelated disorder induces a logarithmic temporal decrease of the defect density, while balanced disorder leads again to an algebraic decay.
\end{abstract}

\maketitle

\section{Introduction}

The nonequilibrium dynamics of quantum many-body systems under the slow change of some external parameter is a challenging and much studied problem. One of the motivations for studying the characteristics of slowly driven systems comes from adiabatic quantum computing \cite{albash,hauke}, a method for solving discrete optimization problems of quadratic, unconstrained, binary type \cite{lucas}. Here, the system is driven from an initial Hamiltonian having an easily preparable, trivial  ground state to a final classical Hamiltonian, typically an Ising model, the ground state of which encodes the solution of the optimization problem \cite{kadowaki,dziarmaga_prl,dziarmaga_random,caneva,dziarmaga_rams,delcampo}. In the aspect of quantum annealing, it is thus desirable to keep the system in its instantaneous ground state, i.e. to drive the system adiabatically so as the final state is the correct ground state giving the solution of the optimization task.  
Otherwise, as it occurs inevitably for finite annealing rates, the final state will be slightly different from the ground state, i.e. it contains some amount of error \cite{kibble,zurek,cz,lukin,king}. According to the adiabatic theorem \cite{amin,albash}, errors form most easily in the presence of low-energy excitations. Therefore, as it is the case for the paradigmatic Ising model driven by a transverse field, the breaking of adiabaticity  occurs essentially when passing through a quantum critical point (QCP) hallmarked by a vanishing energy gap and manifesting in finite systems as an avoided level crossing.   
For slow changes driven through or to a QCP, there is a universal, heuristic scaling theory by Kibble and Zurek which provides scaling relations for the characteristics of the non-adiabaticity of the process such as the error density $n$ in terms of the annealing rate $1/\tau$ \cite{kibble,zurek,cz}. In general, these are power-law relations, like $n\sim \tau^{-\frac{\nu}{1+\nu z}}$,  where the exponents are expressed with the critical exponents of the QCP: the dynamical exponent $z$, describing the vanishing of the energy gap with the system size, $\epsilon\sim L^{-z}$ and the correlation-length exponent $\nu$, characterizing the divergence of the correlation length $\xi$ with the reduced control parameter as $\xi\sim |\Delta|^{-\nu}$. 
This picture is darkened by the circumstance that real optimization problems can usually be represented by Ising models with inhomogeneous couplings. For the transverse-field Ising model with random couplings (or fields), the energy gap at the QCP is known by the strong-disorder renormalization group (SDRG) approach \cite{mdh,fisher,im} to close more rapidly than a power law, as $\epsilon\sim e^{-const\cdot L^{\psi}}$, with $\psi=1/2$ in one dimension, meaning that the dynamical exponent is formally infinite \cite{fisher,young}. This leads to an unfavorable, slower-than-algebraic decrease of the error density with the annealing time: $n\sim(\ln\tau)^{-2}$ \cite{dziarmaga_random,caneva}. 
Thus, disorder leads to an extra deterioration of the accuracy compared to that of the clean system. 
One can then ask the question whether the disorder present in the system through the inhomogeneous couplings given by the optimization problem could be compensated or at least lessened by appropriately chosen inhomogeneous driving fields, so that the closing of the gap could be mitigated. 
For the one-dimensional transverse-field Ising model, the answer is positive.
Here, using local fields combined from neighboring couplings,
it is possible to make the disorder fluctuations independent from the size and to put the scaling of the critical gap back into the realm of power laws \cite{binosi,hoyos_epl,getelina,gh,shirai,juhasz2022}. In this case, we speak of balanced disorder.
We mention that, as it was done in Ref. \cite{knysh}, balancing of coupling disorder can also be achieved by using homogeneous fields but embedding the original chain into a longer chain in which the original spins (logical qubits) are represented by a set of strongly coupled ancillary spins with suitably chosen internal couplings. Nevertheless, by applying the SDRG method to merge ancillary spins, one ends up with an effective model with inhomogeneous fields also in this case. After a series of numerical and analytical work on the random transverse-field Ising chain (RTFIC) with balanced disorder \cite{binosi,hoyos_epl,getelina,gh,shirai}, the dynamical exponent has been recently fixed by using exact lower and upper bounds on the gap \cite{juhasz2022}. 
According to this, the dynamical exponent does not differ from that of the clean system ($z=1$) for sufficiently regular distributions of the couplings, but for distributions having a heavy tail at zero coupling, an anomalous behavior may emerge with $z>1$. Moreover, at the border of normal and anomalous regime, a logarithmic correction also appears.
For a further possibility of reducing the error of the final state by driving a transverse-field with a spatial ramp profile along the chain with a constant velocity, we refer the reader to Ref. \cite{gradient}.

In this paper, we aim at studying the scaling of defect production in two kinds of annealing procedures of the transverse-field Ising chain by Kibble-Zurek theory and numerical calculations.
We will demonstrate in both cases that the temporal scaling of defect density can be made algebraic by balancing the coupling disorder with inhomogeneous fields. One of these processes is a quantum annealing of the closed RTFIC with unitary time evolution, in which the system is driven through its quantum critical point by varying the relative strength of fields and couplings.
Note that, in Ref. \cite{knysh}, a related study, in which balancing was realized by embedding, the coupling distributions were bounded away from zero, so that the QCP always remained in the universality class of the clean system ($z=1$).
In this paper, we will also explore the anomalous regime ($z>1$), as well as the borderline case by using distributions with a power-law tail at zero coupling.
In the second part of this work, we will consider an open variant of the RTFIC. Here, the dynamics is governed by a Lindblad equation which describes a direct coupling of the normal modes of the system to a thermal bath.  
Recently, an environment-temperature quench was studied in the clean variant of this model and a generalization of Kibble-Zurek theory was formulated in Refs. \cite{bd} and \cite{king2023}. We will consider a temperature quench in this model with balanced and also unbalanced disorder, when the system is driven to the QCP by slowly ramping down the temperature to zero, and study the temporal scaling of defect density.

The paper is organized as follows. In Sec. \ref{model}, the model is introduced and its critical properties with balanced disorder are discussed. In Sec. \ref{QA}, technical details of dynamical calculations are presented and the quantities of interest are defined. This is then followed by the formulation of Kibble-Zurek scaling theory and by the presentation of numerical results.
Scaling and numerical results for the environment-temperature quench are presented in Sec. \ref{cooling}. Finally, the results are discussed in Sec. \ref{discussion}.


\section{RTFIC with balanced disorder}
\label{model}

We consider the time-dependent transverse-field Ising chain with open boundary conditions, defined by the Hamiltonian:
\be
{\cal H}(t) =
-\frac{t}{\tau}\sum_{n=1}^{L-1}\frac{J_{n}}{2}\sigma_n^x \sigma_{n+1}^x-\left(1-\frac{t}{\tau}\right)\sum_{n=1}^L\frac{h_n}{2}\sigma_n^z,
\label{H}
\ee
where $\sigma_n^{x,z}$ are Pauli operators at site $n$. In the annealing procedure, the time $t$ is varied from zero to $\tau$ called as the annealing time.
First, we review the static (time-independent) properties of the model, and regard $t/\tau$ as a control parameter of the QCP.   
The solution of the eigenproblem of the Hamiltonian with general couplings $J_n$ and transverse fields $h_n$ is well-known \cite{pfeuty,lsm}.
First, a Jordan-Wigner transformation
\beqn
c_n^{\dagger}+c_n&=&(\prod_{m<n}-\sigma_m^z)\sigma_n^x, \nonumber \\ 
c_n^{\dagger}-c_n&=&i(\prod_{m<n}-\sigma_m^z)\sigma_n^y, \qquad n=1,2,\dots,L \nonumber
\label{JW}
\eeqn
is performed to bring the Hamiltonian in Eq. (\ref{H}) to a quadratic form in fermionic creation and annihilation operators, $c_n^{\dagger}$ and $c_n$, respectively: 
\be
{\cal H}   =
-\sum_{n=1}^{L-1}\frac{J_n(t)}{2}(c_n^{\dagger}-c_n)(c_{n+1}^{\dagger}+c_{n+1})
-\sum_{n=1}^Lh_n(t)(c_n^{\dagger}c_n-\frac{1}{2}),
\label{quad}
\ee
where the notations
\be
J_n(t)\equiv\frac{t}{\tau}J_n, \qquad  h_n(t)\equiv\left(1-\frac{t}{\tau}\right)h_n
\nonumber
\ee
are used.
Then, this Hamiltonian can be diagonalized by a subsequent
Bogoliubov-Valatin transformation, i.e. by introducing new fermionic operators
\be
\eta_k=\sum_{n=1}^L\left[\frac{\phi_{nk}+\psi_{nk}}{2}c_n + \frac{\phi_{nk}-\psi_{nk}}{2}c_n^{\dagger}\right],  \quad k=1,\dots,L
\ee
resulting in
\be
{\cal H}=\sum_{k=1}^{L}\epsilon_k(\eta_k^{\dagger}\eta_k-\frac{1}{2}).
\label{diag}
\ee   
The (real) coefficients $\phi$ and $\psi$ of the transformation, as well as the excitation energies $\epsilon_k$ can be found by performing a singular-value decomposition
\be
M=\phi\mathcal{D}\psi^T
\label{svd}
\ee
of the bidiagonal matrix 
\be
M=
-\begin{bmatrix}
h_1(t)   & &               &        &               &     \cr
J_1(t)   & h_2(t)            &  &        &               &     \cr
    & J_2(t)          &  \ddots           &  &         &     \cr  
    &               & \ddots         &      &         &     \cr
    &               &               &   J_{L-2}(t)   &    h_{L-1}(t)    &  \cr  
    &               &               &        &  J_{L-1}(t)          & h_L(t) 
\end{bmatrix},
\label{M}
\ee
where $\mathcal{D}={\rm diag}\{\epsilon_1,\epsilon_2,\dots,\epsilon_L\}$.

To apply Kibble-Zurek scaling theory (see later), it is necessary to know the finite-size scaling of the energy gap at the critical point. As the dynamics preserves the fermion number parity (which is the eigenvalue of the parity operator $\prod_n\sigma_n^z$), the energy gap accessible by the dynamics is $\epsilon_1+\epsilon_2$.
This, nevertheless, follows in general the same finite-size scaling at the critical point as $\epsilon_1$.

In the case when both $J_n$ and $h_n$ are independent random variables, and the criticality condition $\overline{\ln J(t)}=\overline{\ln h(t)}$ is fulfilled, where the overbar denotes an average over disorder, the lowest excitation energy
scales with the system size as
\be
\epsilon_1\sim e^{-C\sqrt{L}},
\ee
where $C$ is an $O(1)$ (sample-dependent) random variable, as found by the SDRG method \cite{fisher} and confirmed numerically \cite{yr}.   

In the present work, we will focus on balanced disorder which means that the couplings are independent, identically distributed (i.i.d.) random variables, whereas the transverse fields are correlated with neighboring couplings, most generally in the form
\be
h_n=J_{n-1}^sJ_n^{1-s}
\label{balanced}
\ee
for bulk sites, where $s$ is a free parameter in the range $0\le s\le 1$ \cite{shirai}.
With this choice, the critical point of the model is at $t/\tau=1/2$, and the fluctuations of the sample-dependent control parameter $\Delta_L=\sum_{n=1}^{L-1}\ln(J_n/h_n)$ are independent of $L$, as opposed to $\Delta_L\sim \sqrt{L}$ valid for uncorrelated disorder by central limit theorem.

In what follows, we fix the distribution of couplings to a power-law form in the range $(0,1)$ as 
\be
f(J)=\frac{1}{D}J^{-1+\frac{1}{D}},
\label{dist}
\ee
where $0<D<\infty$ is a strength of disorder.
The free parameter $s$ is fixed to $s=0$, so that
$h_n=J_n$ for $1\le n<L$, while $h_L$ was drawn from the same distribution as the couplings. 

In Ref. \cite{juhasz2022}, exact lower and upper bounds on the lowest excitation energy have been established in terms of sums of independent random variables related to the couplings. These imply that, concerning the finite-size scaling of the excitation energy $\epsilon_1$, two regimes can be distinguished depending on the strength of disorder. Specially for $s=0$, balanced disorder is irrelevant for $D<1/2$, in the sense that the scaling of the homogeneous system, $\epsilon_1\sim L^{-1}$ holds to be valid. For $D>1/2$, however, the gap is determined essentially by the smallest coupling present in the system, and we have $\epsilon_1\sim L^{-\frac{1}{2}-D}$. At the borderline case, $D=1/2$, balanced disorder is marginal and merely induces a logarithmic correction of the form $\epsilon_1\sim\frac{1}{L\sqrt{\ln L}}$.
Thus, the dynamical exponent of the model is finite and depends on the strength of disorder as
\be
z=\frac{1}{2} + \max\{D,\frac{1}{2}\}.
\label{z}
\ee

The other ingredient to Kibble-Zurek scaling theory is the correlation-length exponent. According to numerical results on the bulk correlation function in Ref. \cite{hoyos_epl}, the correlation-length exponent of the homogeneous model $\nu=1$ remains unaltered under balanced disorder. 
Now, we justify this finding by making use of the closed form of the surface magnetization at site $n=1$, which is valid for a fixed boundary condition $h_L=0$ at the opposite end of the chain \cite{peschel,ir98}:
\be
\langle\sigma_1^x\rangle=\left\{1+\sum_{l=1}^{L-1}\prod_{n=1}^l\left[\frac{h_n(t)}{J_n(t)}\right]^2\right\}^{-1/2}.
\label{peschel}
\ee
This quantity can be regarded as an end-to-end correlation function $\langle\sigma_1^x\sigma_L^x\rangle$ of the order parameter since, due to the boundary condition, the local order parameter at $n=L$ is fixed to $\langle\sigma_L^x\rangle=1$.  
Evaluating the formula in Eq. (\ref{peschel}) with balanced disorder in Eq. (\ref{balanced}), choosing $h_1=J_1^{1-s}$ for a general $s$, we obtain
\be
\langle\sigma_1^x\rangle=\left(1+\sum_{n=1}^{L-1}e^{2n\Delta}J_n^{-2s}\right)^{-1/2},
\label{m1}
\ee
where
\be
\Delta=\ln(\tau/t-1)
\label{red}
\ee
is a reduced control parameter (defined for $0<t<\tau$).
We can see from this formula that, close to the critical point, $|\Delta|\ll 1$, the magnetization $\langle\sigma_1^x\rangle$ shows essentially critical behavior for small enough system sizes fulfilling $2L|\Delta|\ll 1$, and deviations from criticality appear only well beyond the size $L^*\sim |2\Delta|^{-1}$. This characteristic size can be identified with the correlation length, so we conclude that the correlation-length exponent is $\nu=1$ just as for the homogeneous system, in agreement with numerical results of Ref. \cite{hoyos_epl}.

\section{Quantum annealing with balanced disorder}
\label{QA}

\subsection{Dynamics}

The unitary time evolution of the model during annealing can be conveniently studied by the time-dependent Bogoliubov theory \cite{dziarmaga_random,caneva},
which we will formulate in terms of Clifford operators, $\hat{a}_n\equiv\hat{d}_{n}=c_n^{\dagger}+c_n$ and $\hat{b}_n\equiv\hat{d}_{L+n}=c_n^{\dagger}-c_n$, $n=1,2,\dots,L$.
Using this set of operators, the Hamiltonian in Eq. (\ref{quad}) can be rewritten as
\be
{\cal H} =-\frac{1}{4}\sum_{n,m=1}^{2L}\hat{d}_n^{\dagger}H_{nm}\hat{d}_m,
\ee
where
\be
H=
\begin{bmatrix}
0  & M     \cr
M^T    &  0             
\end{bmatrix}
\ee
with the bidiagonal matrix $M$ defined in Eq. (\ref{M}). 
In the Heisenberg picture, the time-evolved Clifford operators, denoted by $\hat{d}_n(t)$, obey the following equations of motion:
\be 
\dot{\hat{d}}_n(t)={\rm i}\sum_{m=1}^{2L}H_{nm}\hat{d}_m(t).
\label{dt}
\ee
At the beginning of the annealing procedure ($t=0$), a Bogoliubov-Valatin transformation is performed, which can be written in terms of Clifford operators in matrix notation as
\be
\begin{bmatrix}
\hat{\alpha}(0)   \cr
\hat{\beta}(0)             
\end{bmatrix}=
\begin{bmatrix}
\phi^T  & 0     \cr
0    &  \psi^T             
\end{bmatrix}
\begin{bmatrix}
\hat{a}(0)   \cr
\hat{b}(0)             
\end{bmatrix},
\ee
where $\hat{\alpha}_n(0)\equiv\eta_n^{\dagger}(0)+\eta_n(0)$ and $\hat{\beta}_n(0)\equiv\eta_n^{\dagger}(0)-\eta_n(0)$.
The evolution equations can then be solved by the ansatz
\be
\hat{d}(t)=\begin{bmatrix}
\hat{a}(t)   \cr
\hat{b}(t)             
\end{bmatrix}
=
\begin{bmatrix}
{\rm Re}\phi(t)  & -i{\rm Im}\phi(t)     \cr
-i{\rm Im}\psi(t)    &  {\rm Re}\psi(t)             
\end{bmatrix}
\begin{bmatrix}
\hat{\alpha}(0)   \cr
\hat{\beta}(0)             
\end{bmatrix}.
\ee
Substituting this into Eqs. (\ref{dt}), we are led to the following differential equations for the time-dependent Bogoliubov coefficients $\phi(t)$ and $\psi(t)$:
\beqn
\dot{\phi}=-iM\psi \nonumber \\
\dot{\psi}=-iM^T\phi.
\label{diff}
\eeqn

\subsection{Measures of defect production}

To quantify the deviation of the final state from the true ground state, we used two quantities.
One of them is the density of defects, defined by
\be
n(t)=\frac{1}{2(L-1)}\sum_{n=1}^{L-1}\langle\Psi_0|[1-\sigma_n^x(t)\sigma_{n+1}^x(t)]|\Psi_0\rangle,
\label{ed}
\ee
where $|\Psi_0\rangle$ denotes the initial state at $t=0$. 
In the case of a perfectly adiabatic driving, for which the system arrives at the ferromagnetically ordered ground state at the end of the annealing procedure ($t=\tau$), the defect density defined in this way is zero.
The spin operator appearing in Eq. (\ref{ed}) can be expressed with Clifford operators as 
$\sigma_n^x\sigma_{n+1}^x=\hat{b}_n\hat{a}_{n+1}$ and its expectation value can be readily evaluated to be $\langle\hat{b}_n\hat{a}_{n+1}\rangle=-[\psi\phi^{\dagger}]_{n,n+1}$. Thus, the defect density is given by the time-dependent Bogoliubov coefficients as
\be
n(t)=\frac{1}{2(L-1)}\sum_{n=1}^{L-1}(1+[\psi(t)\phi^{\dagger}(t)]_{n,n+1}).
\label{ed2}
\ee

In addition to this, we also calculated the residual energy density, which is the difference between the expectation value of the energy density $E(t)=\frac{1}{L}\langle\Psi_0|{\cal H}(t)|\Psi_0\rangle$ and the instantaneous ground-state energy density $E_{\rm gs}(t)$:
\be
E_{\rm res}(t)=E(t)-E_{\rm gs}(t).
\ee
The latter is obtained by diagonalizing the instantaneous Hamiltonian and using  Eq. (\ref{diag}): $E_{\rm gs}(t)=-\frac{1}{2L}\sum_{k=1}^{L}\epsilon_k$, while the former can be written by straightforward calculations in a compact form in terms of the time-dependent Bogoliubov coefficients:
\be
E(t)=-\frac{1}{2L}{\rm Re}{\rm Tr}\{\phi^{\dagger}(t)M\psi(t)\}.
\label{res}
\ee
Just as the defect density, the residual energy density is zero if the annealing procedure is perfectly adiabatic.

\subsection{Kibble-Zurek scaling theory}
\label{kz}

The basic concept of the Kibble-Zurek mechanism of defect formation is the following. The time evolution of the system is essentially adiabatic sufficiently far from the critical point, up to some freezing time $t^*$. Beyond this time, the state remains unchanged (freezes), while, on the other side of the QCP and sufficiently far from it, the dynamics will be again adiabatic.
Thus, the characteristic size of ferromagnetic domains at the end of the annealing procedure, the inverse of which is the defect density, will be determined by  the correlation length $\xi^*$ at the freezing time.
The freezing point is determined by the condition that, here, the time $\tilde t^*\equiv\frac{\tau}{2}-t^*$ remaining to reach the QCP is comparable with the relaxation time which is given by the inverse of the energy gap \cite{dziarmaga_prl,qKZ}:
\be
\tilde t^*\sim \epsilon^{-1}[\Delta(\tilde t^*)].
\ee
According to Eq. (\ref{red}), $\Delta(\tilde t)\simeq 4\frac{\tilde t}{\tau}$ close to the QCP, i.e. for $|\Delta|\ll 1$.
Using $\xi\sim |\Delta|^{-\nu}$, which is valid for $L\gg\xi$, and the dynamical relationship between the gap and the correlation length  
\be 
\epsilon^{-1}\sim\begin{cases}
\xi & {\rm if} \quad D<\frac{1}{2} \\
\xi\sqrt{\ln\xi} & {\rm if} \quad D=\frac{1}{2} \\
\xi^{\frac{1}{2}+D} & {\rm if} \quad D>\frac{1}{2}, 
\end{cases}
\ee
which follows from the finite-size scaling presented in Sec. \ref{model},
we obtain that freezing occurs for $D\neq\frac{1}{2}$ at
$\tilde t^*\sim \tau^{\frac{z\nu}{1+z\nu}}$ with $\nu=1$ and $z$ given in Eq. (\ref{z}), whereas, for $D=\frac{1}{2}$, at $\tilde t^*\sim \tau^{\frac{1}{2}}(\ln\tau)^{\frac{1}{4}}$.
For the error density in the final state, which is
$n(\tau)\sim 1/\xi^*\sim \Delta(\tilde t^*)\sim \tilde t^*/\tau$, we obtain ultimately
\be
n(\tau)\sim\begin{cases}
 \tau^{-\frac{1}{2}} & {\rm if} \quad D<\frac{1}{2} \\
 (\frac{\tau}{\sqrt{\ln\tau}})^{-\frac{1}{2}} & {\rm if} \quad D=\frac{1}{2} \\
\tau^{-\frac{1}{3/2+D}} & {\rm if} \quad D>\frac{1}{2}. 
\end{cases}
\label{ed_scaling}
\ee
All these relations are expected to be valid asymptotically, for large annealing times $\tau$ and for large enough system sizes fulfilling $L\gg \xi^*\sim 1/n(\tau)$, for which finite-size effects are negligible. 

Next, we consider the residual energy density at the end of the annealing procedure. For $t=\tau$, the instantaneous Hamiltonian is that of a classical Ising model, the excitations of which are domain walls. These domain walls, the density of which is $n(\tau)$, tend to form at links with weak couplings. Thus, we may assume that the fraction $n(\tau)$ of links having the couplings in the range $[0,n^D(\tau)]$ are all excited, giving a mean residual energy density
\be
\overline{E}_{\rm res}(\tau)=\overline{J}_{n(\tau)}n(\tau),
\ee
where $\overline{J}_{n(\tau)}$ denotes the mean coupling under the condition that $J<n^D(\tau)$:
\be
\overline{J}_{n(\tau)}=\frac{1}{n(\tau)}\int_0^{n^D(\tau)}Jf(J)dJ=\frac{D}{1+D}n^D(\tau).
\ee
Here, we made use of the form of the coupling distribution given in Eq. (\ref{dist}). 
We have thus 
\be
\overline{E}_{\rm res}(\tau)=\frac{D}{1+D}[n(\tau)]^{1+D},
\ee
yielding the following asymptotic scaling form of the residual energy density with the annealing time:
\be
\overline{E}_{\rm res}(\tau)\sim\begin{cases}
 \tau^{-\frac{1+D}{2}} & {\rm if} \quad D<\frac{1}{2} \\
 (\frac{\tau}{\sqrt{\ln\tau}})^{-\frac{3}{4}} & {\rm if} \quad D=\frac{1}{2} \\
\tau^{-\frac{1+D}{3/2+D}} & {\rm if} \quad D>\frac{1}{2}. 
\end{cases}
\label{res_scaling}
\ee

\subsection{Numerical results}
\label{num}

In order to check the validity of the scaling relations of the defect density and residual energy density obtained in the previous section, we studied the annealing procedure numerically.
The first step is to obtain the initial values of Bogoliubov coefficients $\phi(0)$ and $\psi(0)$. Since the Hamiltonian in Eq. (\ref{H}) at $t=0$ is already diagonal, these will be simply $\phi_{nk}(0)=-\psi_{nk}(0)=\delta_{nk}$. Then, the differential equations for the Bogoliubov coefficients in Eq. (\ref{diff}) are solved numerically up to time $\tau$, which has been done by the fourth-order Runge-Kutta method. In the final state, we evaluated the defect density and the residual energy density using Eqs. (\ref{ed2}) and (\ref{res}).    
These computations were performed for different system sizes $L=128,256$ and $512$, for different annealing times $\tau=2^n$, $n=2,3,\dots,9$, and for different strengths of the disorder $D=1/3,1/2,$ and $1$. In each case, the computations were performed for $100$ random samples and averages of $n(\tau)$ and $E_{\rm res}(\tau)$ were calculated. 
\begin{figure}[ht]
\includegraphics[width=8cm]{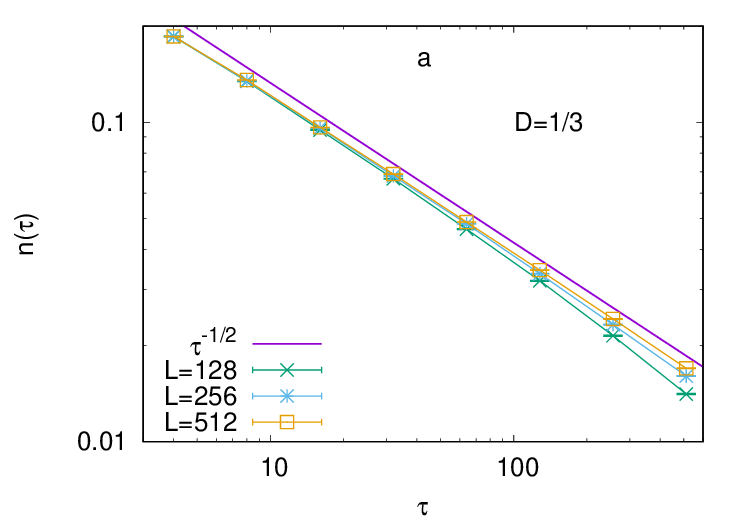}
\includegraphics[width=8cm]{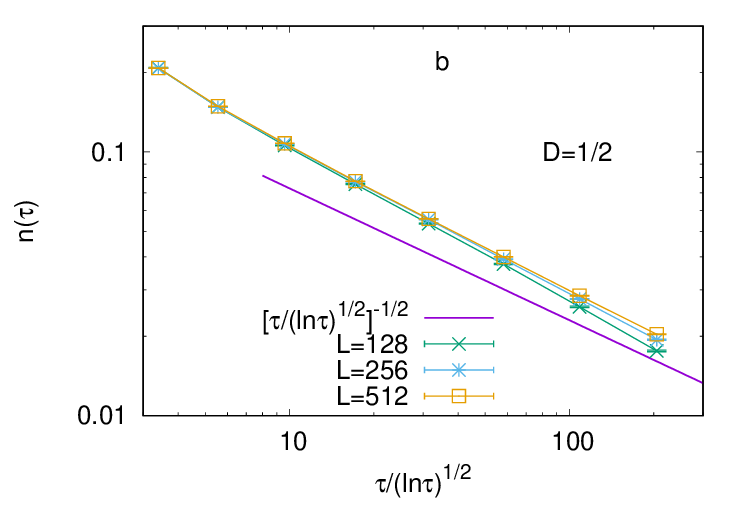}
\includegraphics[width=8cm]{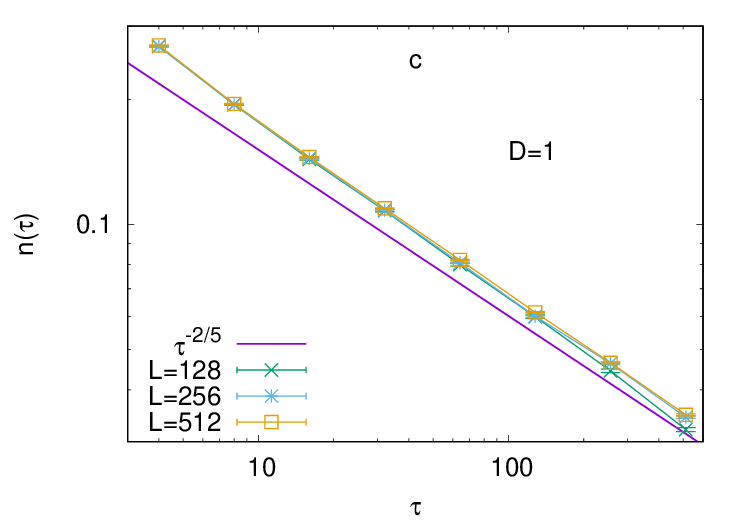}
\caption{\label{fig_ed} The average error density at the end of the annealing procedure plotted against the annealing time for different sizes $L$, for $D=1/3$ (a), $D=1/2$ (b), and $D=1$ (c). The solid lines indicate the predictions of the Kibble-Zurek theory given in Eq. (\ref{ed_scaling}). Power-law fits to the data (in terms of $\tau/(\ln\tau)^{1/2}$ for $D=1/2$) in the large-$\tau$ domain give the following estimates for the exponents: $0.51$ ($D=1/3$), $0.53$ ($D=1/2$), and $0.41$ ($D=1$).
}
\end{figure}
\begin{figure}[ht]
\includegraphics[width=8cm]{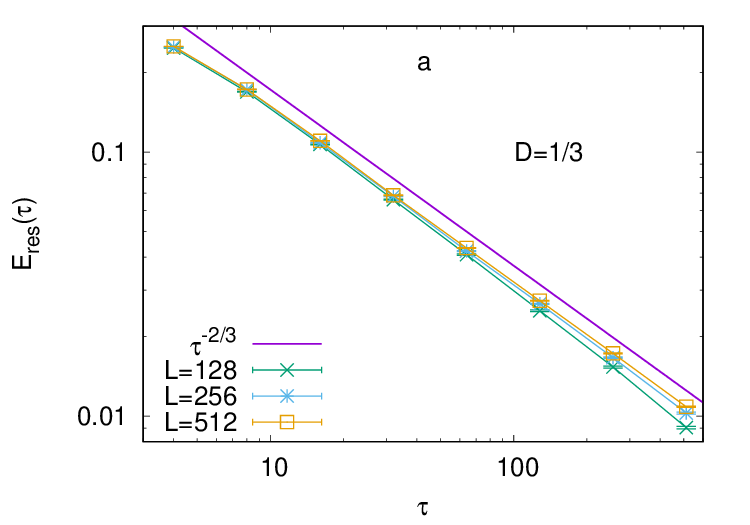}
\includegraphics[width=8cm]{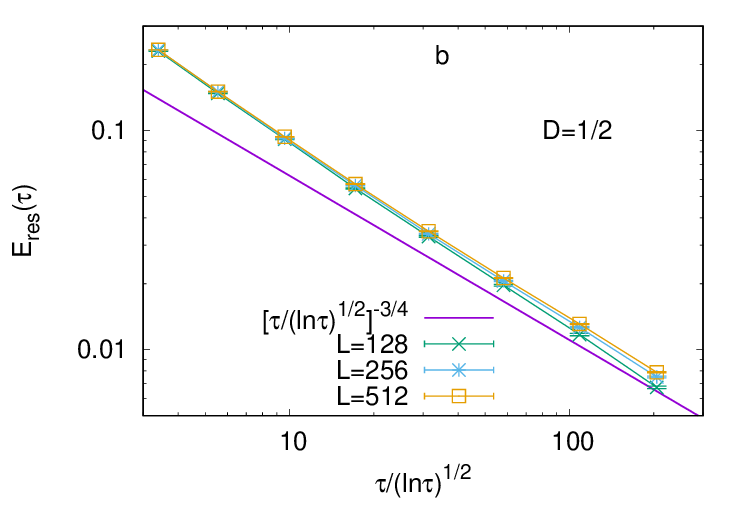}
\includegraphics[width=8cm]{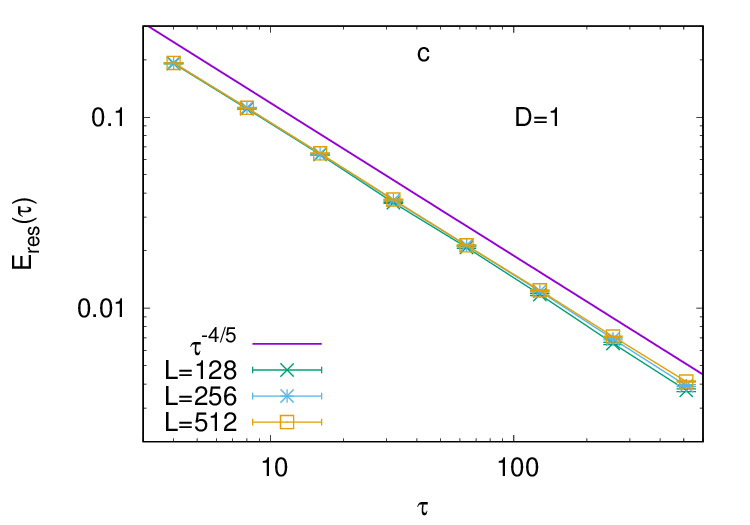}
\caption{\label{fig_res}
The average residual energy density at the end of the annealing procedure plotted against the annealing time for different sizes $L$, for $D=1/3$ (a), $D=1/2$ (b), and $D=1$ (c). The solid lines indicate the predictions of the Kibble-Zurek theory given in Eq. (\ref{res_scaling}).
Power-law fits to the data (in terms of $\tau/(\ln\tau)^{1/2}$ for $D=1/2$) in the large-$\tau$ domain give the following estimates for the exponents: $0.67$ ($D=1/3$), $0.78$ ($D=1/2$), and $0.80$ ($D=1$).  
}
\end{figure}

The average error densities and residual energy densities in the final state are plotted against the annealing time $\tau$ in Figs. \ref{fig_ed} and \ref{fig_res}. As can be seen, the dependence on $\tau$ is in good agreement with the predictions of Kibble-Zurek scaling theory both in the regime, where balanced disorder is irrelevant ($D=1/3$) and in the anomalous regime, where the exponents are different from those of the homogeneous system ($D=1$). At the border of these regimes ($D=1/2$), the corrections to scaling at small $\tau$ are somewhat stronger than in the other two cases, nevertheless, the large-$\tau$ behavior seems to be compatible with the asymptotic law predicted by the theory.

\section{Environment-temperature quench}
\label{cooling}

In the remaining part of the paper, we study an environment-temperature quench of an open variant of the RTFIC, when the system is initially in a thermal equilibrium state, and it is slowly cooled down to zero temperature. 
In an infinitely slow process, the system would remain in thermal equilibrium states throughout the process and would reach its ground state (which is nondegenerate for this model). However, for a finite cooling rate, the final state will be a mixed state with a finite density of defects.
In this process, the Hamiltonian itself is constant and it is chosen to be at its critical point at zero temperature.
We are interested in the question how the density of defects depends on the cooling rate.

\subsection{The model}

To model the environment, we use a Lindblad description, similar to that used in Ref. \cite{bd} for the homogeneous system. The time evolution of the density operator $\rho$ is governed by the equation
\begin{widetext}
\begin{equation}
	\frac{d \rho }{d t} = -i[{\cal H},\rho] + \sum_{n=1}^L \sum_{s=\pm1}\gamma_{n,s}\left( L_{n,s} \rho L^{\dagger}_{n,s} - \frac{1}{2} \left\{ L_{n,s}^\dagger L_{n,s}, \rho \right\}\right). 
\end{equation}
\end{widetext}
Here, the Hamiltonian ${\cal H}\equiv{\cal H}(\frac{\tau}{2})$ of the system is time independent and it is given in Eq. (\ref{H});
the Lindblad operators $L_{n,s}$ are chosen to be the fermionic operators which diagonalize the Hamiltonian, see Eq. (\ref{diag}), as:
\begin{align}
L_{n,+1}&=\eta^{\dagger}_n, \nonumber \\
L_{n,-1}&=\eta_n,
\label{L}
\end{align}
while the coupling constants are 
\be
\gamma_{n,s}=\gamma\frac{1}{1+e^{s\beta \epsilon_n}},  
\ee
with a constant transition rate $\gamma$. Note that these coupling constants fulfill detailed balance at the temperature $T=1/\beta$ of the bath.
For the role played by the coupling to the bath in nonequilibrium scaling behavior in general, we refer the reader to Ref. \cite{yin}. 

Starting from a thermal equilibrium state of the system at some initial temperature, $T_0=1/\beta_0$, we will consider a slow cooling of the environment, i.e. the temperature $T(t)$ is time dependent and ramped down slowly to zero.
Due to that the jump operators are just the diagonal fermion operators, the Linblad equation will decompose into independent dynamical equations for each mode. 
For a general form of time dependence of the temperature, it has been shown in Ref. \cite{bd}, that the occupation of mode $k$ at time $t$ can be obtained as 
\be
  p_k(t)\equiv \langle \eta^{\dagger}_k  \eta_k \rangle = \frac{e^{-\gamma t}}{1+e^{\beta_0 \epsilon_k}} + \gamma \int_{0}^t \frac{e^{-\gamma(t-t')}}{1+e^{\beta(t') \epsilon_k}} d t'.
\label{eq:filling}
\ee

We considered two functional forms for the time dependence of the temperature. One of them is a linear cooling,
\be
T(t)=T_0\left(1-\frac{t}{\tau}\right),
\ee
which is commonly used in the literature. Here, time varies between $t=0$ and $t=\tau$, $\tau$ defining the timescale of the process.
In the case of linear cooling, one has to resort to a numerical evaluation of the integral in Eq. (\ref{eq:filling}). 
The other protocol we considered was the ``hyperbolic'' cooling, given by
\be 
T(t)=\frac{T_0}{1+t/t_0},
\ee
where $t_0$ is a constant of time dimension. Here, the process starts at $t=0$ and can take any length of time.
The hyperbolic cooling is used less frequently, but it has the advantage that one can find a closed formula for the occupation numbers.  Here, the integral appearing on the r.h.s. of Eq. (\ref{eq:filling}) can be shown to be expressed by a hypergeometric function after expanding the integrand in a Taylor series and integrating by terms as: 
\begin{align}
I_k(t) &=  e^{-\gamma t}\int_{0}^t \frac{e^{\gamma t'}}{1+e^{\beta_0\epsilon_k(1+t'/t_0)} } dt' \nonumber \\
&= e^{-\gamma t} \left[e^{\gamma t}  \,_2F_1\left(1,b; c; -e^{-\beta_0\epsilon_k(1+\frac{t}{t_0})}\right)\right]_{0}^{t},
\label{eq:int_hyperbola}
\end{align}
where $b=c-1=-\frac{\gamma t_0}{\beta_0\epsilon_k}+1$.
At late times, i.e. for $\beta_0\epsilon_k\frac{t}{t_0}\gg 1$, one obtains the following asymptotic time dependence of the occupation of modes (with small excitation energies fulfilling $\beta_0\epsilon_k<\gamma t_0$):
\be
p_k(t)=\frac{\gamma t_0}{\gamma t_0-\beta_0\epsilon_k}e^{-\beta_0\epsilon_k(1+\frac{t}{t_0})} + O(e^{-2\beta_0\epsilon_k\frac{t}{t_0}})+O(e^{-\gamma t}).
\label{eq:filling_hyp_long_t}
\ee

In Ref. \cite{chandran2012}, the scaling of the defect density has been investigated for the two types of cooling protocols by a scaling theory and adiabatic perturbation theory for generic quantum and classical systems. As pointed out, the defect density obeys the same scaling law in terms of $\tau$ for linear cooling as in terms of $t/t_0$ for hyperbolic cooling.

To quantify the deviation of the actual state at time $t$ from the (zero temperature) ground state of the system, we considered the number density of fermion excitations
\be
n_f(t)=\frac{1}{L}\sum_{k=1}^Lp_k(t), 
\label{nf}
\ee
and the thermodynamic entropy density of the system given by 
\be
s(t)=-\frac{1}{L}\sum_{k=1}^L\left\{p_k(t)\ln p_k(t)+[1-p_k(t)]\ln[1-p_k(t)]\right\}.
\label{entropy}
\ee

\subsection{Scaling of defect density}
\label{sub:scaling}

Next, we will determine the asymptotic scaling of the number density of fermions, $n_f$, either with $\tau$ for linear cooling or with the time $t$ for hyperbolic cooling. To do so, we need to know the form of the energy density at low energies.

In the case of uncorrelated disorder, i.e. for i.i.d. couplings and either constant or i.i.d. transverse fields, the cumulative distribution of states displays the well-known universal Dyson singularity \cite{dyson1953,eggarter1978} at zero energy: 
\begin{equation}
N_< (\epsilon) \sim \frac{1}{(\ln \epsilon)^2} \;.
\end{equation}
Assuming that the excitation energies are independent random variables, then, according to extreme-value statistics, the typical value of the lowest excitation energy $\epsilon_1$ in a system of size $L$ can be obtained from the condition
\be
LN_<(\epsilon_1)=O(1).
\label{evs}
\ee
This leads to $|\ln\epsilon_1|\sim \sqrt{L}$, which is in agreement with the results of the SDRG method \cite{fisher,young}. Note that, applying Eq. (\ref{evs}) naively to the homogeneous, critical transverse-field Ising chain which has an asymptotically linear dispersion relation, $N_<(\epsilon)\sim\epsilon$, leads again to the correct finite-size scaling of the gap, $\epsilon_1\sim L^{-1}$.  
Concerning the model with balanced disorder of the form $h_n=J_n$, the low-energy tail of the density of states is not known directly; nevertheless, the finite-size scaling of the gap, as presented in Sec. \ref{model}, is available. We can then, heuristically, suppose that the small excitations are independent from each other, and ask what would be the density of states with the known extreme values \cite{note1}.   
Answering this inverse question for balanced disorder yields the following for the distribution of states at low energies:
\begin{align}
N_<(\epsilon) &\sim \epsilon^{1/z} \quad\textrm{if} \quad D \neq 1/2\;, \nonumber \\
N_<(\epsilon) &\sim \epsilon|\ln\epsilon|^{1/2} \quad\textrm{if} \quad D=1/2\;,
\end{align}  
with the dynamical exponent $z$ given in Eq. (\ref{z}).
The density of defects defined in Eq. (\ref{nf}) can be written in the limit of infinite system size as    
\be
n_f(t) = \int p(\epsilon,t) \rho_s(\epsilon)d\epsilon,
\label{nf_int}
\ee
where $\rho_s(\epsilon)=\frac{d}{d\epsilon}N_<(\epsilon)$ is the density of states.

\subsubsection{Linear cooling}

First, let us consider the case of linear cooling. 
Finding the saddle point of the integrand in Eq. (\ref{eq:filling}), one can show that the second term on the r.h.s. at $t=\tau$ is upper bounded by ${\rm const}\cdot \gamma\tau e^{-2\sqrt{\gamma\tau\beta_0\epsilon_k}}$.
This suggests that the occupation number is dominated by the second term of Eq. (\ref{eq:filling}) for long cooling times, $\tau\gg \beta_0\epsilon_k/\gamma$. Moreover, even the contribution of this term is negligibly small if $\gamma\tau\beta_0\epsilon_k\gg 1$. This implies that the upper limit of the integral in Eq. (\ref{nf_int}) can be cut off at $\epsilon=(\gamma\tau\beta_0)^{-1}$ and we may write
\be
n_f(\tau)\sim N_<[(\gamma\tau\beta_0)^{-1}].
\ee
This yields
\be 
n_f(\tau)\sim [\ln(\gamma\tau\beta_0)]^{-2}
\label{nftau_uc}
\ee
for uncorrelated disorder and
\be
n_f(\tau)\sim\begin{cases}
 \tau^{-1} & {\rm if} \quad D<\frac{1}{2} \\
 \frac{\sqrt{\ln(\gamma\tau\beta_0})}{\tau} & {\rm if} \quad D=\frac{1}{2} \\
\tau^{-\frac{1}{1/2+D}} & {\rm if} \quad D>\frac{1}{2} 
\end{cases}
\label{lin}
\ee
for balanced disorder.

\subsubsection{Hyperbolic cooling}

For the hyperbolic cooling, we can see in Eq. (\ref{eq:filling_hyp_long_t}) that, at late times $t\gg\gamma^{-1}$, the occupation of modes having an excitation energy much greater than $T_0t_0/t$ is negligible. We may therefore write
\be
n_f(t)\sim N_<\left(\frac{T_0t_0}{t}\right), 
\ee
which results in
\be 
n_f(t)\sim \left(\ln\frac{t}{T_0t_0}\right)^{-2}
\label{nft_uc}
\ee
for uncorrelated disorder and 
\be
n_f(t)\sim\begin{cases}
 t^{-1} & {\rm if} \quad D<\frac{1}{2} \\
 \frac{\sqrt{\ln(\frac{t}{T_0t_0}})}{t} & {\rm if} \quad D=\frac{1}{2} \\
t^{-\frac{1}{1/2+D}} & {\rm if} \quad D>\frac{1}{2} 
\end{cases}
\label{hyp}
\ee
for balanced disorder.

Concerning the entropy density, one can see that the sum appearing in Eq. (\ref{entropy}) is essentially determined by the $O(1)$ contributions of active low-energy modes, similar to the defect density in Eq. (\ref{nf}). Therefore, the entropy density is expected to follow generally the same scaling with time as the defect density:
\be
s(t)\sim n_f(t).
\ee

Results on the dependence of defect density and entropy density on time found in this section for the two protocols and various types of disorder are summarized in Table \ref{table:2}.
\begin{table}[h!]
\centering
\begin{tabular}{|c|c|c|c|c|} 
\hline
& uncorrelated    & balanced & balanced & balanced \\ 
		&    disorder &   $D<1/2$      &   $D=1/2$     &  $D>1/2$  \\
\hline
\hline
linear & $[\ln(\gamma\tau\beta_0)]^{-2}$ & $\tau^{-1}$ &  $\frac{\sqrt{\ln(\gamma \tau \beta_0)}}{\tau}$ & $\tau^{-\frac{1}{1/2+D}}$   \\[1ex]  \hline
hyperbolic & $\left(\ln\frac{t}{T_0t_0}\right)^{-2}$ & $t^{-1}$ & $\frac{\sqrt{\ln(\frac{t}{T_0t_0}})}{t}$ &  $t^{-\frac{1}{1/2+D}}$ \\[1ex]
		\hline
\end{tabular}
\caption{ Scaling of the defect density and entropy density for different types of disorder during environment-temperature quench.}
\label{table:2}
\end{table}

\subsection{Numerical results}

We calculated numerically the time dependence of defect density $n_f(t)$ and entropy density $s(t)$ for both cooling protocols. The excitation spectrum was obtained by numerically solving Eq. (\ref{svd}) and, using these excitation energies, the occupation numbers were calculated either by numerically evaluating the integral in Eq. (\ref{eq:filling}) (for linear cooling), or by using Eq. (\ref{eq:int_hyperbola}) (for hyperbolic cooling). This has been carried out for 1000 random samples, and the average defect density and entropy density were calculated.  
For linear cooling, the system size was $L=128$ and $256$; the cooling times were $\tau=2^n$, with $n=4,5,\dots,12$.
For hyperbolic cooling, we performed the computations for $L=128,256$, and $512$. The parameters of the model were fixed in all cases to $T_0=1$, $\gamma=0.1$, and $t_0=1$.

\begin{figure}[ht]
\includegraphics[width=8cm]{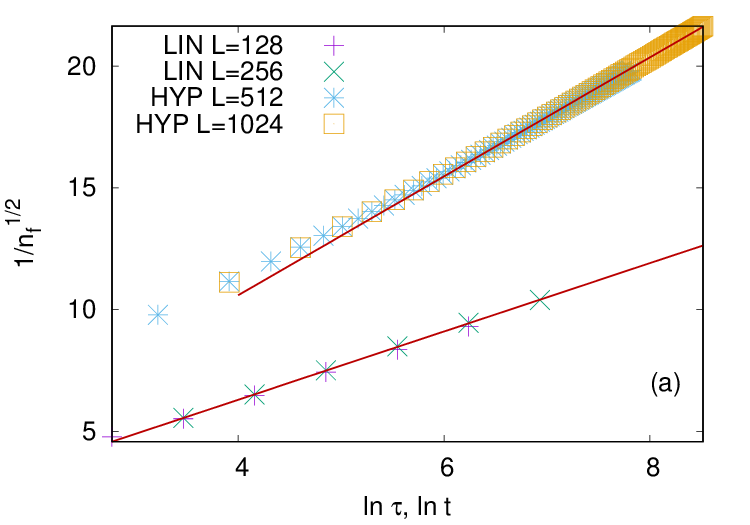}
\includegraphics[width=8cm]{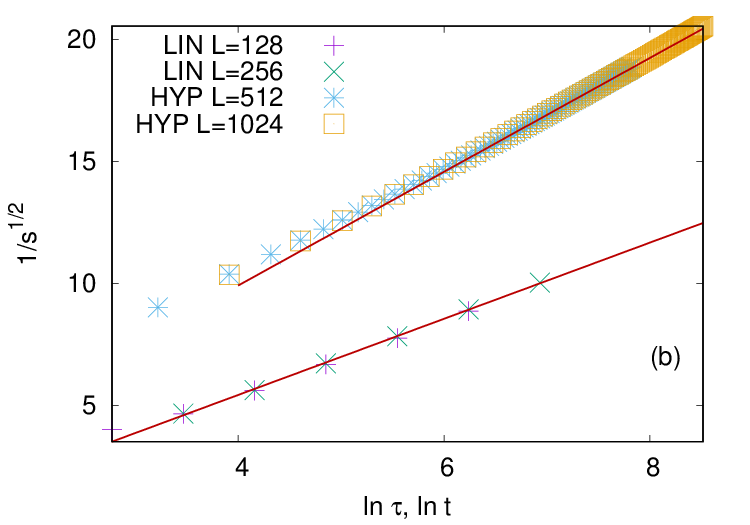}
\caption{\label{fig_cooling_uncorr}
Time dependence of defect density (a) and entropy density (b) under slow cooling in the open RTFIC with uncorrelated disorder.
In each panel, the lower data points correspond to linear cooling (LIN), while the upper curves correspond to hyperbolic cooling (HYP). In the former (latter) case, the horizontal axis is $\ln\tau$ [$\ln t$]. The straight red lines are linear fits to the data for $L=256$ (LIN) and $L=1024$ (HYP).
}
\end{figure}

First, we considered the model with uncorrelated disorder. Here, both the couplings and transverse fields were independently drawn from a uniform distribution given in Eq. (\ref{dist}) with $D=1$. 
The numerical results for the time dependence of defect density and entropy density are shown in Fig. \ref{fig_cooling_uncorr}.
Plotting $n_f(t)^{-1/2}$ and $s(t)^{-1/2}$ against $\ln t$ (with $t=\tau$ for linear cooling), one can indeed see a linear asymptotic dependence at late times, in agreement with Eqs. (\ref{nftau_uc}) and (\ref{nft_uc}). 

\begin{figure}[ht]
\includegraphics[width=8cm]{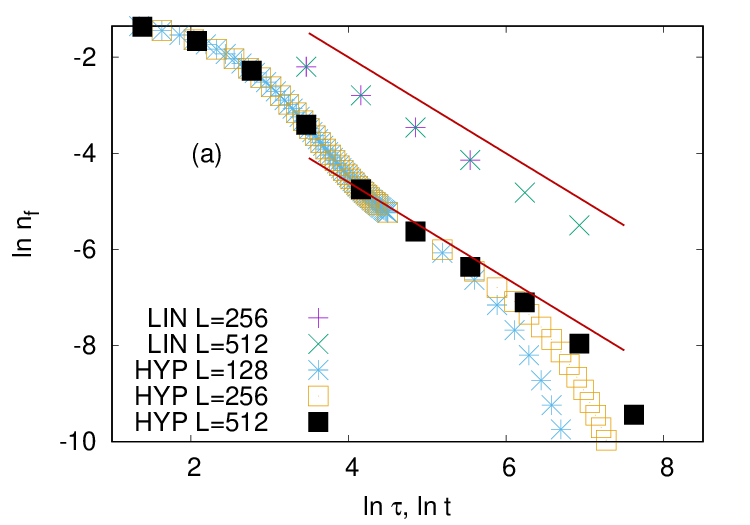}
\includegraphics[width=8cm]{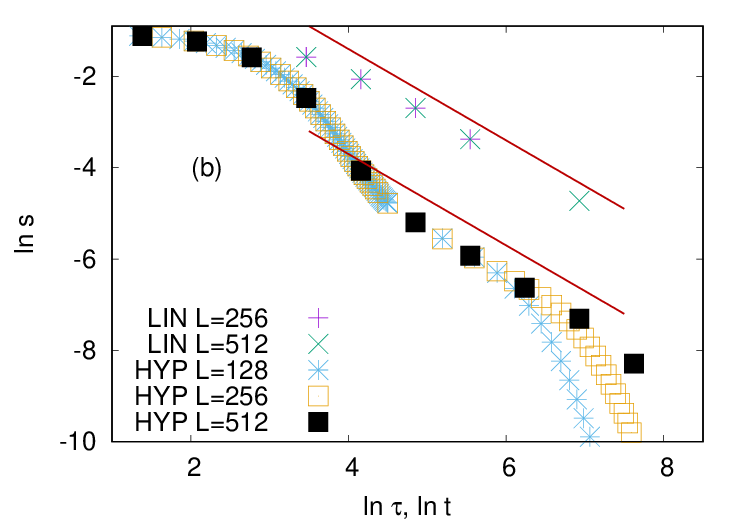}
\caption{\label{fig_cooling_D033}
Time dependence of defect density (a) and entropy density (b) under slow cooling in the open RTFIC using balanced disorder with $D=1/3$.
In each panel, the upper data correspond to linear cooling (LIN), while the lower data correspond to hyperbolic cooling (HYP). In the former (latter) case, the horizontal axis is $\ln\tau$ [$\ln t$]. The straight lines have a slope $-1$. 
Linear fits to the data for the largest system size in the scaling regime give the following estimates of the exponents: $0.98$ (LIN) and $1.06$ (HYP) in the case of the defect density, and $0.98$ (LIN) and $1.03$ (HYP) in the case of the entropy density.
}
\end{figure}

Next, we turn to the model with balanced disorder. Numerical results for $D=1/3$ are shown in Fig. \ref{fig_cooling_D033}. 
In this case, balanced disorder is irrelevant and, according to Eqs. (\ref{nftau_uc}) and (\ref{nft_uc}), both the defect density and the entropy density are expected to decay inversely proportionally to the time. 
As can be seen in the figures, this is indeed the case, after a transient time $\ln t\sim 4$ up to a size-dependent cutoff time, which is, in the case of hyperbolic cooling, at $\ln t\sim 6$ for the largest size. As the limitations of the scaling prediction appear most spectacularly in these data, we briefly discuss the range of validity of scaling results here; nevertheless, the following reasoning is generally valid.
 In the short-time regime ($\ln t<4$ in Fig. \ref{fig_cooling_D033}), the corresponding energy scale $\epsilon\sim t^{-1}$ is not sufficiently low, so that the corrections to the asymptotic behavior of the density of states $N_<(\epsilon)$ used in Sec. \ref{sub:scaling} are still considerable. As a consequence, the temporal scaling of the defect density will deviate significantly from the prediction obtained there.
At late times ($\ln t> 6$ in Fig. \ref{fig_cooling_D033}), the scaling is broken by finite-size effects. The reason for this is that, beyond a cutoff time scale $t\sim 1/\epsilon_1(L)$, which is determined by the energy gap $\epsilon_1(L)$ of the finite system, even the lowest-energy mode will be hardly excited. Therefore, the defect density will quickly decrease here with time; much more rapidly than in the preceding scaling regime.   
Note that, compared to the case of uncorrelated disorder, the finite-size cutoff of the data occurs here earlier, due to the more slowly vanishing gap ($\epsilon_1\sim L^{-1}$).

\begin{figure}[ht]
\includegraphics[width=8cm]{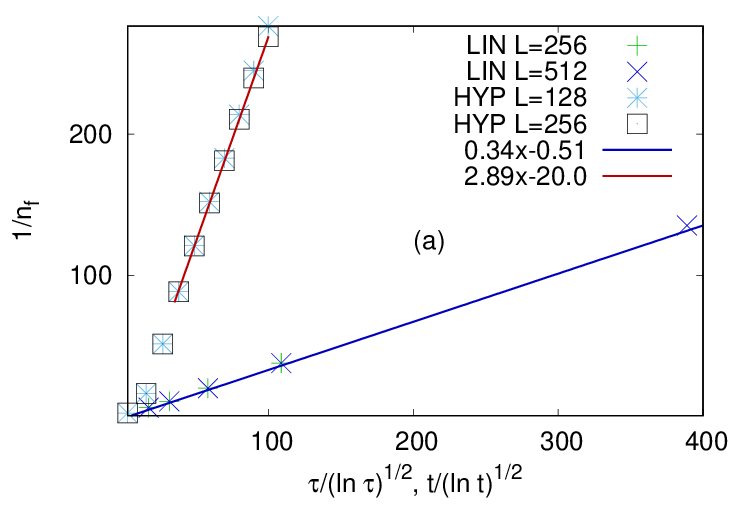}
\includegraphics[width=8cm]{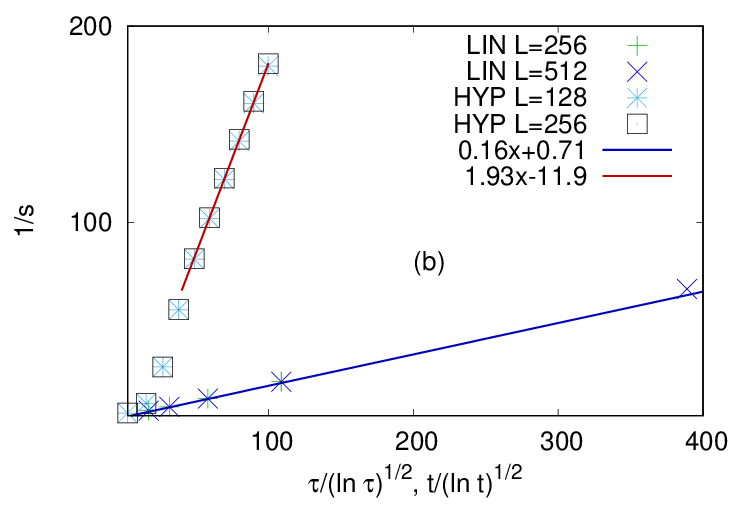}
\caption{\label{fig_cooling_D05}
Time dependence of the inverse of defect density (a) and entropy density (b) under slow cooling in the open RTFIC using balanced disorder with $D=1/2$.
In each panel, the lower data correspond to linear cooling (LIN), while the upper data correspond to hyperbolic cooling (HYP). In the former (latter) case, the horizontal axis is $\tau/\sqrt{\ln\tau}$ [$t/\sqrt{\ln t}$]. The solid lines are linear fits to the data for $L=512$ (LIN) and $L=256$ (HYP).}
\end{figure}

Numerical results obtained in the marginal case $D=1/2$ are shown in Fig. \ref{fig_cooling_D05}. According to Eqs. (\ref{nftau_uc}) and (\ref{nft_uc}), the defect density and entropy density are expected to decrease asymptotically as $n_f(t)\sim s(t)\sim \sqrt{\ln t}/t$ (with $t=\tau$ for linear cooling). As can be seen in the figures, the data are in agreement with this asymptotic form at late times.  

\begin{figure}[ht]
\includegraphics[width=8cm]{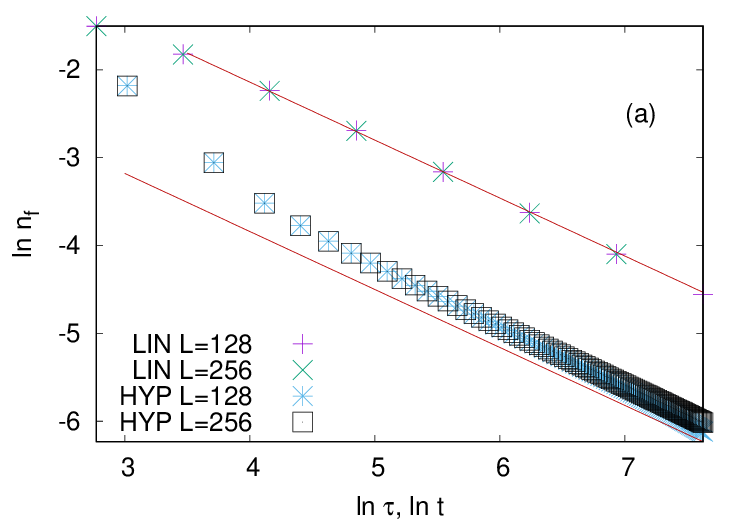}
\includegraphics[width=8cm]{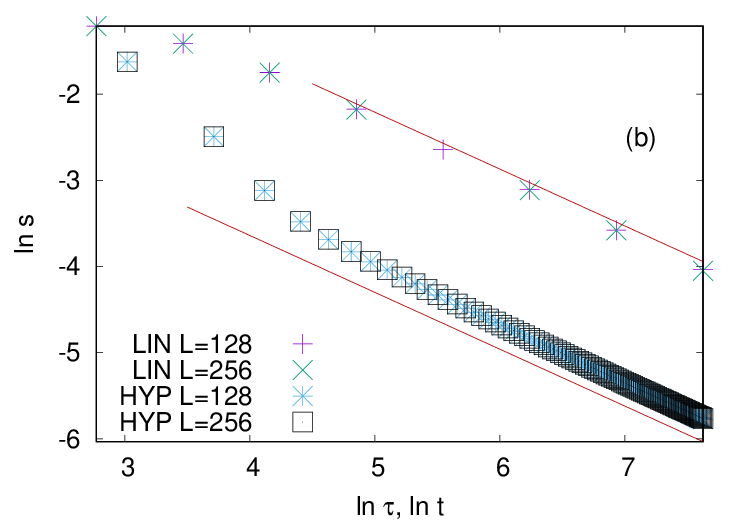}
\caption{\label{fig_cooling_D1}
Time dependence of defect density (a) and entropy density (b) under slow cooling in the open RTFIC using balanced disorder with $D=1$.
In each panel, the upper data correspond to linear cooling (LIN), while the lower data correspond to hyperbolic cooling (HYP). In the former (latter) case, the horizontal axis is $\ln\tau$ [$\ln t$]. The straight lines have a slope $-2/3$.
Linear fits to the data for the largest system size at late times give the following estimates of the exponents: $0.67$ (LIN) and $0.68$ (HYP) for both the defect density and the entropy density. 
}
\end{figure}

Finally, Fig. \ref{fig_cooling_D1} shows the dependence of defect density and entropy density on time for $D=1$. 
According to Eqs. (\ref{nftau_uc}) and (\ref{nft_uc}), these are expected to scale asymptotically as $n_f(t)\sim s(t)\sim t^{-2/3}$. As can be seen in the figures, the numerical data are compatible with this scaling form after a transient time. 

\section{Discussion}
\label{discussion}

In transverse-field Ising models, disorder leads in general to a stretched-exponentially vanishing energy gap at the critical point, and to a slower-than-algebraic (logarithmic) decrease of the defect density with annealing time as the system is driven through its quantum critical point.
For quantum annealing, it is therefore an important question whether the unfavorable effects of disorder can be reduced.   
In this paper, we studied this question in a simple model, the random transverse-field Ising chain.
Considering quantum and thermal annealing in this model, the effects of coupling disorder were balanced by applying appropriately chosen inhomogeneous driving fields which are locally correlated with the couplings.  
In the case of quantum annealing, we adapted the Kibble-Zurek scaling theory of defect formation to this model. For this purpose, we made use of earlier results on the finite-size scaling of the gap \cite{juhasz2022} and, by analyzing the surface magnetization, we justified earlier numerical observations on the robustness of the correlation-length exponent \cite{hoyos_epl}.
Predictions of Kibble-Zurek theory on the scaling of defect density and residual energy density have been confirmed by numerical simulations of quantum annealing. The results consistently demonstrate that both quantities decrease algebraically with the annealing time as opposed to the logarithmically slow decrease in the model with unbalanced random couplings.
For coupling distributions the support of which is bounded away from zero,
the scaling exponents will be the same as in the homogeneous model. Thus, in this case, inhomogeneous fields can completely balance the disorder of couplings.
For distributions having a power-law tail at zero coupling, like that in Eq. (\ref{dist}), the scaling exponents depend on the strength of disorder $D$.
For weak enough disorder, i.e. $D<D_c=\frac{1}{2}$, the scaling exponents characterizing the error production during the annealing process hold to be the same as in the homogeneous system. Thus, balancing with inhomogeneous fields is still perfect.
For $D>D_c$, however, the scaling exponents start to deviate from that of the homogeneous system, and vary with $D$. In this region, the defect density decreases with annealing time slower than in the homogeneous case. Thus, a complete balancing with inhomogeneous fields cannot be realized here.
Although the results of this paper have been obtained for the special choice of $s=0$ in Eq. (\ref{balanced}), the generalization to any value of $s$ is straightforward. For instance, using the general form of the dynamical exponent $z=\max\{Ds,\frac{1}{2}\}+\max\{D(1-s),\frac{1}{2}\}$ from Ref. \cite{juhasz2022}, we obtain that the boundary separating perfect balancing from partial one lies at $D_c(s)=D_c(1-s)=\frac{1}{2(1-s)}$ for $0\le s\le\frac{1}{2}$. 

We also studied thermal annealing in the open variant of the critical RTFIC with uncorrelated, as well as balanced disorder. Here, the environment is slowly cooled down to zero temperature. According to our scaling and numerical results for balanced disorder, the defect density and the entropy density decrease algebraically with the cooling time, with the same exponent, irrespective of the type of cooling protocol (linear or hyperbolic). This latter finding is in agreement with results of Ref. \cite{chandran2012}.
In general, the defect density at the same timescale is found to be smaller for the hyperbolic cooling than for the linear one, and, in the former case, the finite-size effects are found to be stronger.
The algebraic scaling found for balanced disorder is in agreement with the general scaling theory of Refs. \cite{bd} and \cite{king2023} formulated for a finite dynamical exponent. The logarithmic scaling found for uncorrelated disorder is a generalization of that to the case of a Dyson type of singularity in the density of states.

We have seen that the logarithmic scaling of the defect density can be changed to a more favorable algebraic scaling by using inhomogeneous driving fields in any case in the one-dimensional transverse-field Ising model.  
Based on these findings, it is natural to ask whether, in higher-dimensional variants of the model or in the presence of random longitudinal fields, the scaling of defect density could be improved (and if yes, to which extent) by applying optimally chosen inhomogeneous driving fields. These much more difficult questions are left for future research.

\begin{acknowledgments}
This work was supported by the National Research, Development and Innovation Office NKFIH under Grant No. K128989 and by the Quantum Information National Laboratory of Hungary.
\end{acknowledgments}



\begin{thebibliography}{99}
\bibitem{note1}
In the strong-disorder regime $z>1$, the energy gap scales with $L$ in the same way as the smallest coupling present in the system. This suggests that the low excitation energies are essentially given by the weak couplings, justifying thereby the assumption on independence, at least in the $z>1$ regime.
\end{thebibliography}
\end{document}